\documentclass[twocolumn,showpacs,preprintnumbers,amsmath,amssymb]{revtex4}
\usepackage{graphicx}
\usepackage{epstopdf}
\usepackage{bm}

\begin{document}

\title{Entropic determination of the phase transition in a coevolving opinion-formation model}

\author{E.Burgos$^{1}$, Laura Hern\'andez$^{2}$, H. Ceva$^{1}$, R.P.J.Perazzo$^{3}$}
\email{burgos@tandar.cnea.gov.ar, Laura.Hernandez@u-cergy.fr, ceva@cnea.gov.ar, rperazzo@itba.edu.ar}
\affiliation{\small$^{1}$ Departamento de F{\'{\i}}sica, Comisi{\'o}n Nacional de Energ{\'\i }a At{\'o}mica, 
Avenida del Libertador 8250, 1429 Buenos Aires, Argentina \\
$^{2}$ Laboratoire de Physique Th\'eorique et Mod\'elisation, UMR-8089 CNRS-Universit\'e Cergy Pontoise, France\\
 $^{3}$ Departamento de Investigaci\'on y Desarrollo, Instituto Tecnol\'ogico de Buenos Aires, Avda. Madero 399, 
(C1106AD) Buenos Aires, Argentina.}

\date{\today}

\begin{abstract}

We study an opinion formation model by the means of a co-evolving complex network where the vertices represent the individuals, characterised by their evolving opinions, and the edges represent the interactions among them. The 
network  adapts to the spreading of opinions in two ways: not only connected agents  interact and eventually change 
their thinking but an agent  may also rewire one of its links  to a neighborhood holding the same opinion as his. The dynamics depends on  an external parameter $\Phi$, which controls the plasticity of the network. 
We show how the information entropy associated to the distribution of group sizes, allows to locate the phase transition between full consensus and a society where different opinions coexist. We also determine the minimum size of the most informative sampling. At the transition the distribution of the sizes of groups holding the same opinion is  scale free.
\end{abstract}
\maketitle

\section {Introduction}

The behaviour of large communities of individuals, may be studied using the concepts and methods of statistical physics~\cite{fortunato}.  We follow this line to consider the case of the build up of opinion groups in a population. Opinions spread within a population 
via person-to-person contacts where they are subject to controversy and discussion. Any 
two agents holding  different opinions may, after being in contact,  either keep their previous opinions  or change them  and eventually, coincide. 
In this process agents with the same thinking may become more numerous
constituting large opinion groups while opinions held by few agents may lose relevance and eventually, dissapear.

The social changes involved in the spread of opinions and the formation of opinion groups can be studied by mapping 
this problem into the evolution of a social graph in which each node represents an agent characterized by  a  variable representing its opinion,  while the links represent the contacts (interactions) among the agents.  

Several works that consider the situation of binary opinions use the framework of the Ising model for magnetic 
materials (e.g \cite{Nowak}, \cite{binar1}, \cite{binar2}). The case of three-opinon states has also been studied, mapping it into a Blume-Emery-Griffith model~\cite{BEM}. The term {\em sociophyisics} was coined to refer to these kind of models.   On the other  extreme, as may be  the case of religious beliefs, opinions may actually be of an infinite 
variety and can be assimilated to a continuous variable. The {\em bounded confidence model} considers the situation where the interaction between two agents depend on how similar their opinions already are~\cite{deffuant}  
An intermediate situation is considered in the  CODA (continuous opinion, discrete actions) model~\cite{martins} which interpolates between discrete actions, taken at some stage of the dynamics, based on evolving continuous  opinion variables.  Another  intermediate situation, appears when  several discrete opinions are possible, as  for an election with many candidates. In this case the opinions are described by an discrete variable (see e.g. ref \cite{HyN}).

While people exchange opinions on a personal basis, it is also true that agents that agree, naturally tend to 
gather in closer communities while those with different opinions segregate. In studying the formation of these  groups, the problem is what comes first: either  
opinions spread over  the topology of the network forming clusters  of agreeing individuals or a change in the topology brings together agents having the same opinion that were not in contact before.  Recently  this point has been  studied introducing  the coevolution of nodes and links. Two 
mechanisms that mutually interfere with each other are considered: one is the change of the individual opinions by the successive 
interactions with other agents and the other is the  change in the structure of the neighborhood of each agent thus 
conditioning its possible interactions~\cite{HyN,Laurapaper}. The co-evolution of both adaptation mechanisms may be controlled  by an external parameter, as in~\cite{HyN} where 
 a change in the opinion is produced with probability $1-\Phi$, ($\Phi \in [0,1]$),  while the topology of the 
network is changed with probability $\Phi$. Alternatively, the co-evolution of nodes and links may depend on a dynamical variable as in~\cite{cultdyn} or finally, both dynamics may be independent~\cite{Laurapaper}.

In ref.~\cite{HyN} an adaptation algorithm is proposed such that for extreme values of $\Phi $ either the system evolves 
toward a state with a single large group of agents sharing one opinion, or to a distribution 
of groups of agents with different opinions. The remarkable result is that for some intermediate value of $\Phi$ 
a dynamic phase transition is found  accompanied by, among other things, a power-law distribution of group sizes.
Despite some similarities of this transition with the random graph percolation case, these authors prove that 
it belongs to a different universality class. 

In ref.~\cite{marsili} it is shown how  the behavior of a complex system 
 may be studied using the tools of information theory, namely by measuring  the entropy associated to the probability distributions with which different states of the system appear when  sampling it under  some given conditions. This sampling is in general incomplete and it is shown that one can identify the  maximally informative samples of such systems, which show a power law distribution of relevant quantities.

In the present work we apply this latter approach to an opinion formation model that we have chosen, for comparison, similar (though not identical) to the one studied in ref.~\cite{HyN}. The main goal of this approach is to characterize the relevance of different samplings in terms of the amount of information that  they give about the behavior of a complex system.

A pertinent choice of variables allows us to locate the phase transition using the entropy associated to the distribution of such variables. This method also allows us to determine the minimum size of the most   informative sample.

\section{The model}  

We consider a society of $N$ agents, each one having an opinion that is labeled by an integer variable $\omega_i = 1 \dots \Omega$.
 
No metric is assigned to the opinion labels. We describe the community as a graph in which each node represents an 
individual. The interaction among agents is only allowed when the corresponding nodes are neighbors, 
i.e. they are joined by a link of the graph. The total number $M$ of links of the graph, as well as the opinions, are initially distributed at random. 

Starting from that initial configuration, the social graph is allowed to evolve. Such evolution takes place in discrete 
time steps in which links and opinions are assumed to co-evolve. In each time step, as in ref.~\cite{HyN} either an opinion is changed with 
probability $1-\Phi$ or a link is changed with probability $\Phi$, nevertheless the dynamics is different.

At each step a node is chosen at random representing the {\it active} agent. 
Then one randomly chooses one of its neighbors among those holding a different opinion. With probability $1-\Phi$, 
the active agent  
 confronts its opinion with the chosen neighbor. The result of such interaction is obtained applying a \textit{ global majority rule} by which the 
node  holding the opinion with less supporters of the two, adopts that of its counterpart, instead of simply copying its neighbor's opinion as in~\cite{HyN} . With probability 
$\Phi$  rewiring takes place. This means that  the   
 link joining the active agent to the chosen neighbor   is cut   and  the active agent is reconnected to any other agent of the system that holds its  opinion.

According to this algorithm, in each step either an opinion or a link, is changed. When $\Phi \approx 0$ opinions are changed very often
and the topology of the network remains essentially unchanged while if $\Phi \approx 1$, the opposite happens: opinions are left 
unchanged but the topology of the graph is modified.  This procedure keeps the total number of links, $M$,  constant. In either case 
links between agents having different opinions are gradually eliminated and replaced by links between agents with the same thinking.
The adaptation process therefore converges to a situation in which there are no links between agents with different opinions. 
  The main difference between this scheme and  the one proposed in ref \cite{HyN}  is the use of a global  majority rule, which  appears here in both stages of the dynamics. This represents the discussion between the two agents who finally choose to agree on the best accepted opinion of the two, instead of one agent simply copying the opinion of the other. This also differs from the {\em local} majority rule, where the active agent is only influenced by its nearest neighbors.  As a  result,   not only  convergence is significantly improved but also the  critical value of the rewiring parameter, $\Phi_c$, is shift.

During the adaptation process, opinion groups may change size by either growing or dwindling, causing eventually 
some opinions to disappear. Once the number of links between agents with different opinions vanishes, the social graph remains 
segmented into a set of disconnected subgraphs each one with agents of a single opinion. This does not mean that each opinion 
is represented by a connected graph: agents with a same opinion may occupy the nodes of several disconnected subgraphs.

Once the convergence of each realization is achieved,  each agent has acquired one opinion $\omega_i \in [1,2,\dots \Omega]$. So the $N$ agents are distributed in several opinion groups with $k(\omega)$ adherents ($0 \leq k(\omega)\leq N$). The opinions that have dissapeared in the final state correspond to  $k(\omega)=0$. So  
$\sum_{\omega=1}^{\Omega}k(\omega)=N$. The number of opinion groups with $k(\omega)=n$ is 

\begin{equation} 
m_n=\sum_{\omega=1}^\Omega \delta_{n,k(\omega)}
\label{mk} 
\end{equation}

Since $\sum_{n=0}^N m_n=\Omega$, we can define the probability of finding a group with $n$ members as $P(m_n)=m_n/\Omega$. Notice that these $m_n$ groups do not necessary hold the same opinion and that $m_0$, the number of groups with no member, counts de number of initial opinions that eventually dissapear in the co-evolution.

 Following the ideas of ref.~\cite{marsili}, we calculate the information entropy contained in the distributions of the variables sampled in this model. Studying the evolution of the corresponding entropy one can determine the most informative sampling.

In the present case, a state of the system is given by the outcome of the adaptation process providing a distribution of  the opinions accross the population. The distribution 
of the probability $P(\omega)$ that a randomly chosen agent has the opinion $\omega$ encodes part of the information that can be extracted by sampling the system. In the large $N$ limit, this is defined by 
\begin{equation}
P(\omega)= \frac{k(\omega)}{N}
\end{equation}
 Correspondingly, it is possible to define
the opinion entropy $S_{\Omega}$ as

\begin{equation}
S_{\Omega}=- \sum_{\omega=1} ^{\Omega} {P(\omega)\log[P(\omega)]}
\end{equation}

 One can also be interested in the  probability that an agent belongs to a group of size $k$:
\begin{equation}
P(k)= \frac{km_k}{N} 
\label{pk}
\end {equation}
with the corresponding information entropy: 
\begin{eqnarray} 
S_k & = & -\sum_{k=1}^{N} P(k)\log [P(k)] =  \nonumber  \\
    & = & -\sum_{k=1}^{N}\frac{km_k}{N}\log [ \frac{km_k}{N} ] = \nonumber  \\ 
    & = & -\sum_{k=1}^{N} \frac{km_k}{N}\log(m_k)-\sum_{k=1}^{N} \frac{km_k}{N}\log[\frac{k}{N}] 
\end{eqnarray}
Replacing the value of $m_k$ from eq.(\ref{mk}) and changing the sums over $k$ to sums over $\omega$ 
one obtains:
\begin{eqnarray}
S_k &=& -\sum_{k=1}^{N} \frac{km_k}{N}\log(m_k)-\sum_{\omega=1}^{\Omega} \frac{k(\omega)}{N}\log[\frac{k(\omega)}{N}] \\
    &=&-\sum_{k=1}^{N} \frac{km_k}{N}\log(m_k) +S_{\Omega} 
    \label{entros}
\end{eqnarray}
thus, $S_k \leq  S_{\Omega}$. 

Within the present model there is some degree of ambiguity concerning how to perform averages when $N_r$ 
realizations are made to obtain statistically significant results. All the above derivations are valid 
for each realization separately, thus if each realization is labeled by $r$, one can write the above equations as:
\begin{eqnarray}
S_{\Omega}(r)&=& -\sum_{\omega=1}^{\Omega}P(\omega,r)\log[P(\omega,r)] \\
S_k(r)       &=& -\sum_{k=1}^{N} \frac{km_k(r)}{N} \log[\frac{km_k(r)}{N}]
\end{eqnarray}
and averages can trivially be defined by
\begin{eqnarray}
\overline{S_{\Omega}}&=& \frac {1}{N_r}\sum_{r} S_{\Omega}(r)\\
\overline{S_k}       &=& \frac {1}{N_r}\sum_{r}S_k(r)
\end{eqnarray}
Average entropies also fulfill $\overline{S_k}< \overline{S_{\Omega}}$.

However there is a second possibility, namely to work with the distribution of group sizes measured over the $N_r$ realizations. Then the average number of groups of a given size is:

\begin{equation}
\overline{m_k}= \frac {1}{N_r}\sum_{r}m_k(r) ,
\end{equation}
 One can  then define the information entropy associated to this global  distribution of group sizes: 
\begin{eqnarray}
S_{<k>}&=&-\sum_{k=1}^{N}\frac{k\overline{m_k}}{N}\log[\frac{k\overline{m_k}}{N}] \\
        &=&-\sum_{k=1}^{N}\frac{k\overline{m_k}}{N}\log(\overline{m_k})+\overline{S_{\Omega}}
\label{entros2}
\end{eqnarray}
In this case $S_{<k>} \neq \overline{S_k}$ showing that the result of  calculating the group entropy of each realization and averaging over the sample (average of group size entropies over the $N_r$ realizations, $\overline{S_k}$), is different  from that of  measuring the group probability  distribution over all the realizations of the sample and calculating the entropy associated  with the distribution so obtained ($S_{<k>}$). In addition the above relationship $\overline{S_k}< \overline{S_{\Omega}}$ is no longer fulfilled if  $\overline{S_k}$ is
replaced by $S_{<k>}$.

\section{Results} 
 We have studied systems of $N=400,800,1600,3200$ and $6400$ agents with a random initial distribution of $M$ links, leading to a  connectivity of average degree $c= 4,8$ and $12$. The total number of initial opinions, $\Omega$, goes from very low values ($\Omega=2$)  to very high values ($\Omega=640$).  Averages are typically taken over $N_r=5000$ realizations, except for the largest sizes or the highest connectivities, where we have averaged over  1000 realisations. In the transition region we have performed up to 10000 realizations for all the sizes. The results presented in this article correspond to the case  $c =4$.

\begin{center}
\begin{figure}[tbp]
\includegraphics[width=9cm]{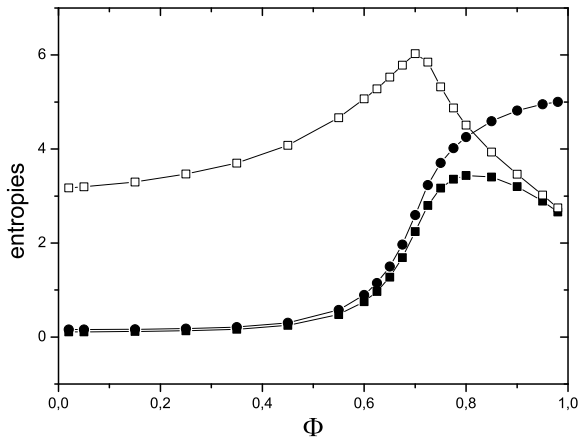}
\caption{Entropies corresponding to the probability distributions of opinions, and sizes of opinions groups.
The social graph has a total of $N=1600$ agents and and $M=3200$ links. They share a total of $\Omega = 160$ opinions.
Averages are made over 5000 realizations. Open boxes correspond to $S_{<k>}$  while  
filled boxes (circles) correspond to $\overline{S_k}$ ($\overline{S_{\Omega}}$) }
\label{entro}
\end{figure}
\end{center}

The topology of the resulting social graph critically depends upon the value of $\Phi$. When $\Phi \approx 0$, 
 opinion changes are enhanced, the social graph approaches a \textit{consensus state} in which a vast majority 
of agents merges into a single giant opinion group. As $\Phi$ grows, a richer spectrum of sizes takes place until a 
moment in which the probability distribution of the sizes of opinion groups approaches a power law. In ref \cite{HyN} this 
situation is assimilated to a dynamical phase transition. For even larger values of $\Phi$ the probability density of  the sizes of opinion groups changes into a bell-type distribution that corresponds to the initial 
random assignment of opinions, due to the fact that for high $\Phi$ rewiring is dominant.    

In Fig. \ref{entro} we show the results  for the average  opinion entropy, $\overline{S_{\Omega}}$, the average of group size entropies $\overline{S_k}$ and the global entropy of  group sizes, $S_{<k>}$ as 
a function of the adaptation parameter $\Phi$.  The plot of $\overline{S_{\Omega}}$ is easy to interpret: when  $\Phi \approx 0$ 
only very few opinions survive because the system approaches the consensus state in which many opinions are left without any 
agents to support them. On the other hand, when $\Phi$ grows the entropy also grows because individual opinions are left 
essentially unchanged with respect to the initial random assignment. 

The same analysis holds for the average entropy of groupe sizes, $\overline{S_k}$, except for the fact that for $\Phi=1$, where only rewiring is possible its value must coincide with $S_{<k>}$. This  can 
be understood if one bears in mind that for this value of $\Phi$, the number of supporters of each opinion is left  unchanged and   therefore  all the opinions are expected to have a  number of followers that fluctuates around  $N/\Omega$, as in the initial distribution. 

\begin{center}
\begin{figure}[tbp]
\includegraphics[width=9cm]{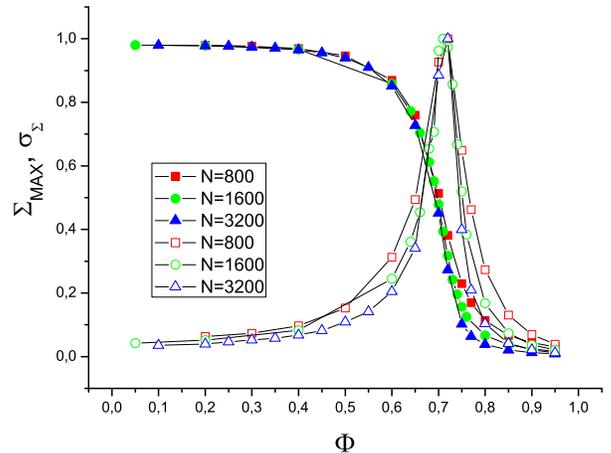}
\caption{Average size (open symbols) and normalised dispersion(full symbols) of the largest cluster as a function of $\Phi$, for different sizes. (color online)}
\label{cluster_max}
\end{figure}
\end{center}

 The curve for  $S_{<k>}$, where the entropy is evaluated using the  group size distributions of all the realizations in the sample, is particularily interesting. Starting at a low value for $\Phi \approx 0$, where there are very few possible group sizes because the system is close to the consensus 
state, it develops sharp maximum  for a particular value of $\Phi_c$. This maximum signals the occurrence of a phase transition between a consensus state and a fragmented one, where  groups holding different opinions coexist.
This is  confirmed by the behaviour of the order parameter $\Sigma_{Max}$, the normalized size of the maximumm cluster which suddenly collapses at $\Phi_c$.
Fig.~\ref{cluster_max}  shows  this collapse of $\Sigma_{Max}$ along with the peak of its dispersion, 
$\sigma_{\Sigma}$, at  $\Phi_c$ respectively for different sizes. It is worthwhile noticing that, besides a slight sharpening of the dispersion curve peak, there is no relevant size effect.

\begin{center}
\begin{figure}[tbp]
\includegraphics[width=9cm]{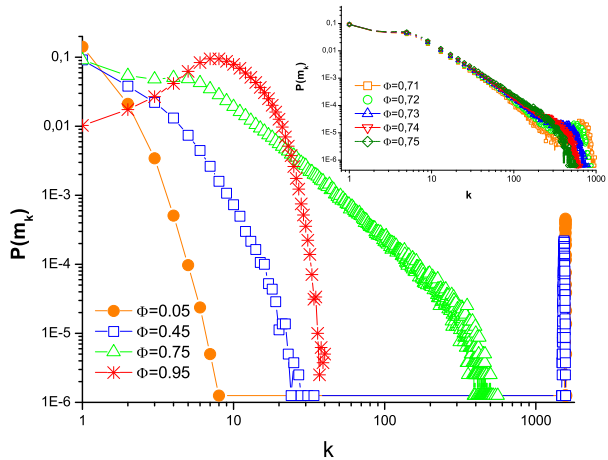}
\caption{Distributions of group sizes  for $N=1600$ agents average degree $c=4$ and  $\Omega = 160$ inital opinions. The results correspond to a sampling of $N_r=10000$ realizations. In the inset, for the sake of clarity the plot shows only one every four measured points. Notice that for $\Phi = 0.71$ a trace of the peak at high k, still remains. (color online)}
\label{disgru}
\end{figure}
\end{center}

In Fig. \ref{disgru} we show several examples of probability distributions of the group size, calculated over all the sample, for different values of $\Phi$. 

For $\Phi \approx 0.05$ the distribution decays fast for very small group sizes
an displays a significant peak for a size of order $N$, showing that in most of the samples  there is a large dominant group, corresponding to the consesus state which may  coexist with some minoritary groups of different opinions. On the other extreme, the distribution for $\Phi \approx 0.95$ corresponds instead, to a bell shape distribution having a maximum located al $k \approx N/\Omega$. 
The inset show different group size distributions in the transition region ($\Phi \in (0.71,0.75)$) where the curve corresponding to the lower bound shows a reminiscence of the peak of the consensus state (large $k$ values).

 In the transition region,  the distribution function of group sizes fits a power law, $P(m_k)\propto k^{-\alpha}$ with $\alpha \approx 2.2$,  which corresponds as expected, to a situation in which group sizes appear with a maximum spread. This dependance has also been observed in ref.~\cite{HyN} but with a larger exponent. The corresponding probability, $P(k)$, that an agent taken at random belongs to a group of size k,  given by eq.~\ref{pk},  also follows a power law but with a different exponent $P(k)\propto k^{-\beta}$, with $\beta \approx 0.95$.

The value of $\Phi_c$ increases with  the connectivity, $c$, of the network,  we have found  $\Phi_c =0.85 $ for $c=8$ and $\Phi_c =0.9$ for $c=12$, indicating that when the connectivity is large, consensus is always reached within this model.  The scale free behaviour of the distribution of group sizes cannot be observed if the number of initial opinions is too low. For very low values of $\Omega$ ($\Omega\approx 10$), and high enough values of $\Phi$, one observes a multi-peak distribution with the same number of peaks as the number of initial opinions.

\begin{center}
\begin{figure}[!ht]
\includegraphics[width=9cm]{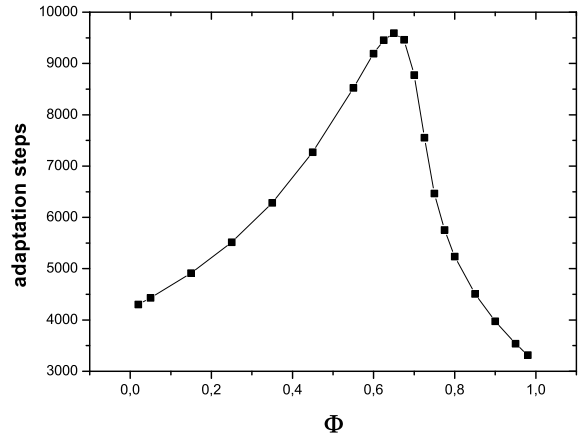}
\caption{Plot of the average number of adaptation steps required to reach a social graph without links between 
agents having different opinions tor  a system with $N=1600$, $\Omega = 160$. Averages 
are made over 5000 realizations. The  maximum is located in the critical region.  }
\label{pasos}
\end{figure}
\end{center}

Interestingly, at $\Phi_c$ the plot of the  number of adaptation steps  required for the social graph to converge to a stationary state presents a well developed a peak, as it is  shown in Figure~\ref{pasos}. Adaptation steps bear a close relationship with computing time, however 
the latter  is not a practical measure because the computing time required by an adpatation step strongly depends upon the value of 
$\Phi$. The existence of this peak in the number of adaptation steps  is consistent with the power law behaviour of the   distribution of group sizes. This is reminiscent of the critical slowing down observed in equilibrium critical phenomena, where the correlation time is related to the divergence of the correlation length, revealing the existence of fluctuations at all scales. Here, instead of domains of all sizes, as in magnetic models,  we have broad distribution of the sizes of groups of  agents holding equal opinions.

\begin{center}
\begin{figure}[!ht]
\includegraphics[width=9cm]{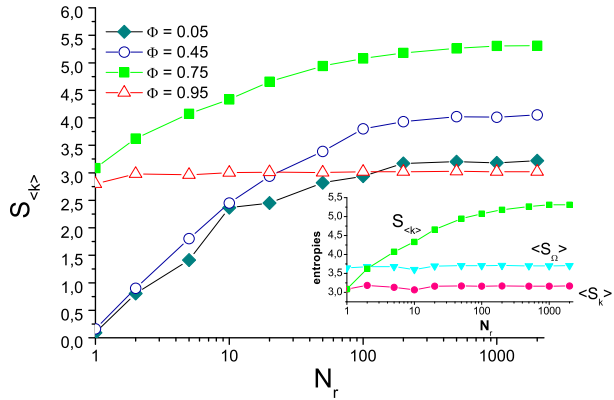}
\caption{Dependence of $S_{<k>}$ on the size of the sampling for different values of $\Phi$ above, below and at the critical region, for a system of $N=1600$  agents and $\Omega=160$ opinions. In the inset, the behaviour of the three entropies defined here, with the size of the sampling in the critical region ($\Phi=0.75$). (color online) }
\label{sampling}
\end{figure}
\end{center}

Figure~\ref{sampling} shows that $S_{\overline k}$ grows until saturation, with the size of the sampling, $N_r$, except for very large 
$\Phi$ values where it remains constant (which correspond to unchanged opinion groups). This allows us to determine the  number of realizations that gives the most informative sampling. Beyond that number, computing more realisations will not bring additional information. As expected ref.~\cite{marsili}, the most informative sampling (the one with the largest entropy), corresponds to the transition region.

 In the inset, the behaviour of the other entropies calculated here, in the critical region, is depicted. 
While $S_{\overline k}$ increases with the size of the sampling until saturation, as described in  the inset of the  Fig. 2 (right) of ref.~\cite{marsili}, the other average entropies  $\overline{S_\Omega}$ and  $\overline{S_k}$  remain constant.  This means that if one calculates the entropy for each realisation and averages over many realisations,  the fact of increasing the sampling size, will not bring any new information.

\begin{center}
\begin{figure}[!ht]
\includegraphics[width=9cm]{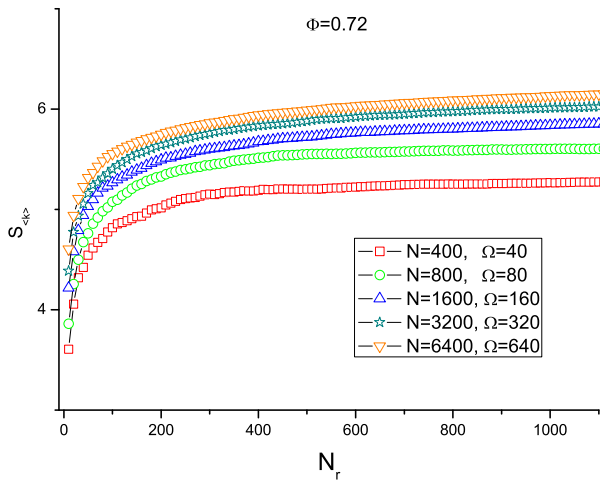}
\caption{ $S_{<k>}$ as a function of the number of realisations $N_r$, for different system sizes in the critical region ($\Phi=0.72$),  all the samples have the same ratio $N / \Omega = 10$. In order to detail the transitory regime the $N_r$ axis shows samplings up to $N_r= 1000$ realisations. We have calcuated  $S_{<k>}$ over samplings containing up to  $N_r= 10000$ realisations and the entropy remains constant. (color online)}
\label{entro_size_eff}
\end{figure}
\end{center}

Interestingly we see that the system is not self averaging. If it were, a sampling consisting of large size networks (large $N$) and few realizations (small $N_r$) should give similar results as those issued from a sampling consisting on a lot of realisations $N_r$ of smaller systems (provided that $N$ is still reasonably big so as  not to be of the same order of magnitude as the number of inital opinions). Figure~\ref{entro_size_eff} shows that this is not the case. The entropy in the critical region is measured for several systems sizes, differing in one order of magnitude, as a function of $N_r$. All the sizes reach saturation at approximately the same value of $N_r$ (obviously the saturation value of $S_{<k>}$  does depend on $N$).

\section{Conclusions}

We perform a study of the phase transition in an opinion model using the information entropy associated to the distribution of different variables. In this model the opinions of the agents and the topology of the social network evolve on the same time scale. This coevolution is controlled by a probabilistic adaptation parameter $\Phi$ that controls the plasticity of the network by allowing to rewire links between agents of different opinions. 

Our results show that the system undergoes a phase transition between a consensus phase and a fragmented one, where several opinions coexist, as was found in ref.~\cite{HyN}, however as  the dynamics used here is different, the critical value of  the rewiring parameter we obtain is different, here: $\Phi_c \approx 0.73$. This can be easily understood: in both  models the imitation probability is $1-\Phi$, but as majority rule is more efficient in creating and sustaining consensus than simply copying the neighbors opinion,  consensus may remain, in this model, for   values of $\Phi$ that are  higher than the critical value    obtained in ref.~\cite{HyN} namely, $\Phi_c \approx 0.46$.  We observe that when  minority rule is applied instead, the convergence is severley hampered and even impeded, leading to a frustrated situation.

We  show that this phase transition may be located using the information entropy associated with the distribution of groups sizes, measured over the sampling of $N_r$ realisations. In the vicinity of  $\Phi_c$ the corresponding distribution of group sizes is scale free,  as expected in the transition region.
We have found that within this model the entropy of the distribution of the sizes of opinion groups is not  self averaging.  In particular, the way in which the average over the different realisations is calculated is far from being irrelevant. This phenomenon is characteristic of disordered magnetic systems, where for instance, the response functions as the specific heat or the susceptibility of each realisation, shows a  well developed peak which it is located at a different temperature for each realisation ~\cite{noself}. Calculating the average of the corresponding the response function over all the realisations first, in order to determine the transition temperature by the location the peak of the average curve, gives a  very broad maximum and the information about the transition is washed out.

As usual when dealing with complex systems, the choice of the variables that we are sampling in order to get the most complete information on the system is crucial. Here the correct sampling is given by $P(\overline m_k)$, and its corresponding information entropy $S_{<k>}$, which improves with the number of realisations, until saturations as it is  shown on  ref~\cite{marsili}. On the contrary, the average  entropy  $\overline S_k$ will not give more information if we increase the number of realisations.  The choice of the order in averaging is relevant. 
Moreover, as  the system is not self averaging, large systems need as many  realisations as small systems in order to reach the most informative regime (where   entropy is maximum).

\textsc{Acknowledgements}.
The authors wish to acknowledge helpful and encouraging discussions with Miguel A. Virasoro.

\end{document}